\documentclass[%
 aip,
 pof,
 amsmath,amssymb,
 preprint,%
]{revtex4-1}

\usepackage{graphicx}
\usepackage{dcolumn}
\usepackage{bm}

\usepackage[utf8]{inputenc}
\usepackage[T1]{fontenc}
\usepackage{mathptmx}
\usepackage{etoolbox}

\usepackage{physics}
\usepackage{dimnum}
\usepackage{mathtools}
\usepackage{verbatim}
\usepackage{subcaption}
\usepackage{gensymb}

\usepackage{color}

\makeatletter
\def\@email#1#2{%
 \endgroup
 \patchcmd{\titleblock@produce}
  {\frontmatter@RRAPformat}
  {\frontmatter@RRAPformat{\produce@RRAP{*#1\href{mailto:#2}{#2}}}\frontmatter@RRAPformat}
  {}{}
}%
\makeatother
\begin{document}

\preprint{AIP/123-QED}

\title{Counterflow around a cylinder}

\author{Matheus P. Severino}
\email{matheus.severino@usp.br}
 
\author{Leandro F. Souza}%
 \email{lefraso@icmc.usp.br}
\affiliation{%
 Instituto de Ciências Matemáticas e de Computação, Universidade de São Paulo \\ 13566-590 São Carlos -- São Paulo, Brazil
}

\author{Elmer M. Gennaro}
\email{elmer.gennaro@unesp.br}
\affiliation{
 Departamento de Engenharia Aeronáutica, Universidade Estadual Paulista \\ 13876-750 São João da Boa Vista, São Paulo, Brazil
}

\author{Daniel Rodríguez}
\email{daniel.rodriguez@upm.es}
\affiliation{%
 Escuela Técnica Superior de Ingeniería Aeronáutica y del Espacio, Universidad Politécnica de Madrid \\ 28040 Madrid, Madrid, Spain
}

\author{Fernando F. Fachini}
\email{fernando.fachini@inpe.br}
\affiliation{
 Grupo de Mecânica de Fluidos Reativos, Instituto Nacional de Pesquisas Espaciais \\ 12630-000 Cachoeira Paulista, São Paulo, Brazil
}

\date{\today}

\begin{abstract}
The incompressible flow around a circular cylinder, positioned at the center of an unconfined planar counterflow, is studied by means of numerical solutions of the conservation equations and linear stability analysis. 
The flow is completely defined by the Reynolds number ($\Rey$) -- based on the cylinder radius, the strain rate defining the counterflow, and the kinematic viscosity. 
For very low values of $\Rey$, the flow is steady, two-dimensional, and fully attached to the cylinder wall. 
Increasing $\Rey$ above $\Rey_s \approx 16.86$, the flow separates, giving rise to two symmetric, counter-rotating recirculation regions on each side of the cylinder. 
Further increasing $\Rey$ leads to a progressive enlargement of the recirculation regions and the appearance of multiple recirculation centers, akin to Moffatt eddies. 
However, the convective acceleration imposed by the counterflow limits their size. 
An oscillatory mode becomes linearly unstable for $\Rey_{c} \approx 4146$. 
This mode gives rise to a sinuous meandering of the wake flow, on each side of the 
cylinder, being analogous to the well-known von K\'arm\'an instability. 
The frequency of this mode is directly proportional to the strain rate defining the counterflow.
\end{abstract}

\maketitle

\section{Introduction}\label{sec:intoduction}

The flow around a cylinder immersed in a uniform stream is a canonical 
configuration in fluid mechanics, acting as a simplified representation 
of bluff-body flows. 
Regardless of its simplicity, it encompasses a wide variety of physical phenomena that have been the subject of comprehensive research 
\citep{williamson/1996:ARFM,buresti/1998}.
In the incompressible limit, the characteristics of the flow are dictated by the Reynolds number ($\Rey$) -- representing the ratio of inertial to viscous forces. 
As it increases, the flow undergoes a sequence of topological and dynamic bifurcations. 
First, the flow separates from the cylinder walls, giving rise to a steady two-dimensional flow with two symmetric recirculation zones \citep{taneda/1956:JPSJ,SEN/2009:JFM}. 
Further increasing $\Rey$ leads to the instability of the steady two-dimensional flow. 
The appearance of a linear instability in the form of a two-dimensional global mode originates a Hopf bifurcation towards a time-periodic two-dimensional flow, characterized by sinuous meandering of the wake and 
later to the shedding of counter-rotating vortices 
\citep{Provansal1987:JFM,jackson/1987:JFM,dusek/1994:JFM}, forming the Mallock--Bénard--von Kármán vortex street 
\citep{mallock/1907:PRSA,benard/1908:CRAS,vonkarman/1911:NMW}.
The shedding frequency is typically described by the \Sr[l] (a dimensionless frequency), which exhibits a well-defined relationship with the \Rey[l] under this regime \citep{Williamson1988:JFM}. 
Even greater increases in the Reynolds number lead to three-dimensional (3D) instabilities within this periodic wake \citep{noack/1994:JFM}, initiating a transitional regime that, ultimately, evolves into turbulence.
The study of wake inception and transition holds significant engineering importance, as they enable the prediction and mitigation of undesirable phenomena, such as flow-induced vibrations in structures 
\citep{williamsoon/2004:ARFM}.

Since then, this canonical configuration has been explored with different bluff bodies, e.g., triangular and elliptical cylinders \citep{jackson/1987:JFM}, flat plates at various angles \citep{jackson/1987:JFM, fage/1927:PRSL}, and square cylinders in normal and rotated incidence \citep{sohankar/1997:JWEIA}.
The mechanism for destabilizing the flow around these bodies is the same as above.
The difference is in the base flow, established by the interaction of the farfield uniform flow and the body.
The main fact is the position of the flow separation because it controls the vorticity production into the shear layer, and the size of the vortices in the recirculation zone \citep{gerrard/1966:JFM}.
For the circular cylinder, the flow separation initiates at the trailing edge and moves towards the equator of the cylinder, by increasing the Reynolds number \cite{taneda/1956:JPSJ}.
The condition of stability for other body configurations changes because the flow separation changes.

The counterflow is a basic configuration for combustion 
\citep{tsuji/1982:PECS}, providing conditions to study flame stability, and allowing the modeling of a local flame element embedded within an arbitrary flow field.
It is a conceptual framework known as the flamelet model \citep{peters/1984:PECS}, which is particularly prominent in turbulent combustion modeling.
Flame stability analysis, in terms of heat loss and flow strain rate, has been extensively performed, as exemplified by the classical asymptotic analysis of flame extinction by \citet{linan/1974:AA}.
On the other hand, the hydrodynamic stability analysis of counterflows is a much more unexplored topic, with few studies for confined domains \citep{pawlowski/2006:JFM}. 

\begin{figure}[t]
\centering
\includegraphics[width=350pt]{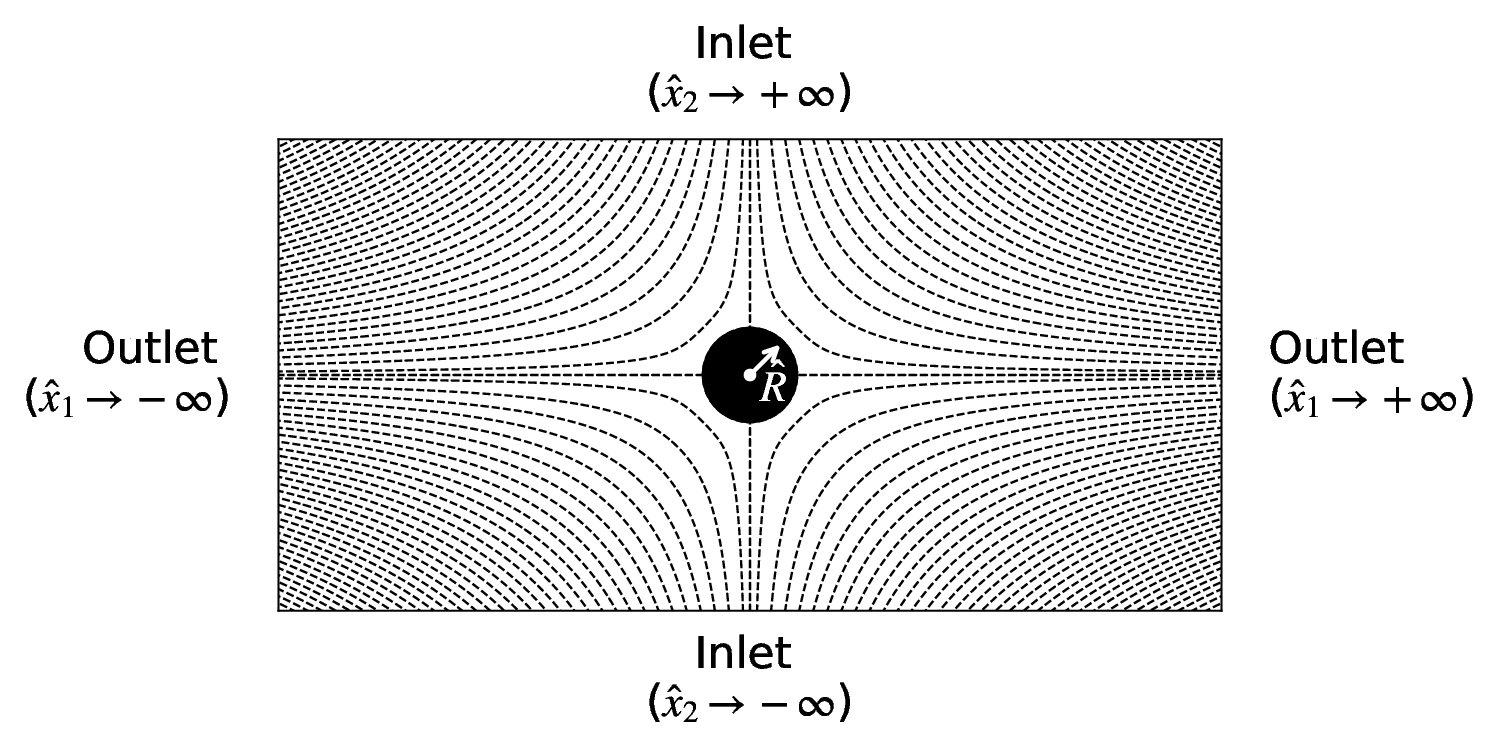}
\caption{
Schematic representation of the flow:
a very long circular cylinder in an unconfined planar counterflow.
}
\label{Fig-scheme}
\end{figure}

This work considers a third fundamental configuration: a combination of the two 
above geometries:
an infinitely long circular cylinder is positioned at the center of an unconfined 
planar counterflow, as depicted in Fig. \ref{Fig-scheme}.
This configuration can be used in heat exchangers to enhance the thermal energy transfer by jet impingement \citep{Zuckerman/2006}.
It is also the basis of the double Tsuji burner \citep{severino/22:AMM, fachini/2025:CST}: 
a configuration introduced to study compact diffusion flames under a gradual change of regime for reactants flow, from counterflow to accelerating coflow \citep{severino/21:CTM}.
The distinction of this setup, in relation to the other bluff-body cases, is the scale of the recirculation zone and, consequently, the vorticity intensity of the 
counter-rotating vortices.
Also, they are still encapsulated in the recirculating zone for a higher Reynolds number because they are under higher pressure, imposed by the counterflow.
  
The existing literature does not investigate this configuration in stability analyses, even at the most basic level.
This work addresses it, avoiding complexities generated by thermal expansion and high-temperature chemical reaction.
Therefore, an incompressible, two-dimensional flow is considered, to focus on the sequence of topology and dynamic bifurcations that take place as the Reynolds number is augmented from the creeping flow limit up to $\Rey = 10\,000$.
Numerical solutions of the conservation equations are used to compute steady flow solutions and characterize the flow topology. 
A linear stability analysis -- based on the calculation of two-dimensional eigenmodes -- is then performed. 
The inherent spatial symmetries of the flow configuration are exploited to reduce the computational cost and guide the analysis. 

The remainder of the analysis is organized as follows: 
Section \ref{sec:matModel} describes the mathematical formulation;
Section \ref{sec:Numerical_methods} briefly covers the numerical methods employed in the 
spatial discretization and computation of the steady flow solutions; 
The linear stability analysis is described in Section \ref{sec:LST};
The results for the steady flow characterization and linear stability analysis are given 
in Section \ref{sec:Results};
Finally, some conclusions are provided in Section \ref{sec:Conclusions}.

\section{Mathematical Formulation}\label{sec:matModel}

Consider a cylinder at the center of an incompressible, two-dimensional counterflow, as shown in Fig. \ref{Fig-scheme}. 
The evolution of this system is described by the momentum conservation and continuity equations
\begin{equation}\label{Eq.CM}
    \partial_t \vb*{u}
    +
    \div
    \left(
        \vb*{u}
        \otimes
        \vb*{u}
    \right)
    =
    -
    \grad
    p
    +
    \dfrac{1}{\Rey}
    \laplacian
    \vb*{u},
\end{equation}
\begin{equation}\label{Eq.C}
    \div \vb*{u} = 0
\end{equation}
in which $\vb*{u} \coloneqq [u_1,u_2]^T$ and $p$ are the  non-dimensional 
velocity and pressure, functions of time $t$ and spatial coordinates 
$\vb*{x} \coloneqq [x_1,x_2]^T$.
The variables are nondimensionalized as
\begin{equation}
        t \coloneqq \hat{t} / \hat{t}_c, \quad \vb*{x} \coloneqq \hat{\vb*{x}}/ \hat{l}_c, \quad
        p \coloneqq \hat{p}/ \hat{p}_c, \quad \vb*{u} \coloneqq \hat{\vb*{u}}/ \hat{u}_c,
\end{equation}
where, $\hat{t}_c \coloneqq \hat{l}_c / \hat{u}_c$ (residence time), 
$\hat{l}_c \coloneqq \hat{R}$ (cylinder radius), 
$\hat{u}_c \coloneqq \hat{R} \hat{a}$ (velocity based on the counterflow 
strain-rate, $\hat{a}$), and $\hat{p}_c \coloneqq \hat{\rho}_c \hat{u}_c^2$ 
(dynamic pressure with constant density $\hat{\rho}_c$), and the Reynolds number is defined as $\Rey \coloneqq \hat{l}_c \hat{u}_c/\hat{\nu}_c$, for a constant kinematic viscosity $\nu_c$.
Superscript $(\ \hat{}\ )$ is used for dimensional quantities.

\subsection{Boundary Conditions}
Equations (\ref{Eq.CM}) and (\ref{Eq.C}) are complemented by the following boundary conditions. In what follows, $\vb*{e}_i$, $\vb*{n}$, and $\vb*{t}$ denote unit vectors in the $x_i$ ($i=1,2$), cylinder outward normal, and tangent directions, respectively, and $\vb*{\sigma} \coloneqq -p\vb*{I} + \Rey^{-1}\left[\nabla\vb*{u} + 
(\nabla\vb*{u})^T\right]$ is the stress tensor. Finally, $d_{x_i}$ is the domain extension in the $i$-direction, defined hereafter.

The potential counterflow solution is imposed at the inlet boundaries,
\begin{equation}\label{eq:in}
\vb*{u} = x_1 \vb*{e}_1 - x_2 \vb*{e}_2, \quad \text{at} \quad x_2 = d_{x_2}.
\end{equation}
Zero-stress boundary conditions are imposed at the outlet
\begin{equation}\label{eq:out}
\vb*{\sigma} \dotproduct \vb*{n} = \vb*{0} \quad \text{at} \quad
x_1 = d_{x_1}.
\end{equation}
No-slip is imposed at the cylinder surface
\begin{equation}\label{eq:cyl}
\vb*{u} = \vb*{0} \quad \text{at} \quad x_1^2 + x_2^2 = 1.
\end{equation}
Finally, symmetry is imposed about the coordinate axes, which takes the form:
\begin{equation}\label{eq:sym}
\vb*{u} \dotproduct \vb*{n} = 0,  \quad
\vb*{t} \dotproduct  \left(\vb*{\sigma} \dotproduct \vb*{n}\right) = 0, \quad \text{at} \quad 
x_1 = 0 \text{  and  } x_2 = 0.
\end{equation}

\subsection{Initial Condition}

The initial condition is based on the potential flow solution, obtained as the superposition of the planar counterflow and a quadrupole at the origin of the spatial coordinates system \citep{severino/21:CTM, severino/22:AMM}:
\begin{equation}
    \vb*{u}(\vb*{x}, t= 0) 
    = 
    \left(
        x_1 \vb*{e}_1 
        - x_2 \vb*{e}_2
    \right)
    - 
    \dfrac{1}{\norm{\vb*{x}}^6} 
    \left[
        x_1(x_1^2 - 3x_2^2) \vb*{e}_1 
        + 
        x_2(3x_1^2 - x_2^2) \vb*{e}_2 
    \right],
\end{equation}
in which $\| \cdot \|$ is the Euclidean norm.

\section{Numerical Methods}
\label{sec:Numerical_methods}

The conservation equations are discretized and integrated using the open-source finite element framework \texttt{Gridap} \citep{badia/2020:JOSS,verdugo/2022:CPC}. The spatial discretization employs the classical and robust (monolithic) mixed Taylor--Hood ($Q_2/Q_1$) elements on structured meshes, utilizing a skew-symmetric convective formulation for discrete kinetic energy conservation. 

Given that strongly convective problems require modifying the symmetric weighting (Galerkin) to prioritize
upstream information \citep{brooks/1982:CMAME}, the present formulation employs a Streamline Upwind and Pressure-Stabilizing Petrov-Galerkin 
method \citep{hughes/1986:CMAME}, and also includes a Grad--Div term (divergence penalization) to enhance local mass conservation, preventing spurious non-solenoidal modes.

\subsection{Domain Size and Grid Structure}

The symmetry-reduced domain $\Omega = [0, 20] \times [0, 10]$ is discretized using a body-fitted multi-block structured mesh, transitioning from orthogonal curvilinear blocks at the wall, to Cartesian-aligned blocks in the wake.
By construction, the mesh has 25--35 elements across the boundary layer for the largest Reynolds number considered ($\Rey = 10\,000$), and over 26 points per wavelength for the highest instability frequencies.
A total of $329\,541$ second-order elements were used, resulting in $1\,302\,771$ nodes, with $2\,598\,323$ degrees of freedom for velocity and $326\,461$, for pressure.

\subsection{Computation of steady-state solutions}

Steady flows are computed using a Newton--Raphson method based on a direct sparse LU decomposition and with trust-region globalization. The Newton--Raphson iteration is considered to be converged when the residual norm falls below $10^{-9}$ and the norm of the solution change between two iterations is below $10^{-11}$. 
For base flows used in the linear stability analysis, the tolerance criteria are refined to $10^{-12}$ for both norms.

\section{Linear Stability Analysis}
\label{sec:LST}

The linear stability is analyzed by studying the temporal evolution of modal disturbances with infinitesimal amplitude superposed to the base flow. The flow-field state $\vb*{q} = (\vb*{u},p)^T$ is decomposed as
\begin{equation} \label{eqn:LST_decomposition}
    \vb*{q}(\vb*{x},t) 
    = 
    \bar{\vb*{q}}(\vb*{x}) 
    + 
    \varepsilon \tilde{\vb*{q}}(\vb*{x}) 
    \exp(\lambda t)
    + 
    \text{c.c.},
\end{equation}
in which, $\bar{\vb*{q}}$ is the steady base flow, $\tilde{\vb*{q}}$ is the spatial eigenvector,  $\lambda = \sigma + i\omega$ is the complex eigenvalue with growth rate $\sigma$ and frequency $\omega$, $\varepsilon \ll 1$ and $c.c.$ denotes the complex conjugate.

Linearizing Eqs. (\ref{Eq.CM}) and (\ref{Eq.C}) about $\bar{\vb*{q}}$, and neglecting higher order terms on $\varepsilon$, yields the PDE-based generalized eigenmode problem
\begin{equation}\label{eqn:LNS_op}
    \lambda \vb*{\mathcal{M}} \tilde{\vb*{q}} = \vb*{\mathcal{L}} \tilde{\vb*{q}}.
\end{equation}
Here, $\mathcal{L}$ is the linearized Navier-Stokes operator and $\mathcal{M}$ is the singular mass operator, defined as $\text{diag}(1, 0)$ for the  $(\tilde{\vb*{u}},\tilde{p})$ variables, reflecting the absence of time derivative in the continuity equation. 
Modes with $\sigma > 0$ have exponential growth in time (unstable), while $\sigma < 0$ indicates exponential decay (stable).
Neutral modes ($\sigma = 0$) neither grow nor decay. 
Consequently, the base flow is linearly stable if all its eigenmodes satisfy $\sigma < 0$.

The same spatial discretization used for the computation of the base flows is applied here. 
The discretized eigenproblem is solved via a matrix-forming approach \citep{gennaro/2013:AIAA,rodriguez/2017:ICOSAHOM} using a shift-and-invert strategy. 
The shifted operator is factorized with direct sparse LU, and eigenvalues are computed using the Krylov–Schur algorithm \citep{stewart/2001:SJMAA}, with a prescribed tolerance of $10^{-9}$.

\subsection{Solution families and boundary conditions}
\label{LSA_familiesAndBC}

The base flow symmetry about $x_1 = 0$ and $x_2 = 0$ allows the linear stability analysis to be restricted to the first quadrant, bounded by $x_1 \in [0,d_1]$ and $x_2 \in [0,d_2]$, without loss of generality \cite{sirovich/1990:PF}.
Modal perturbations are separated into four symmetry families ``XY'', in which X, Y $\in \{\text{S, A}\}$ denote symmetric (S) or antisymmetric (A) behavior with 
respect to $x_2 = 0$ and $x_1 = 0$, respectively \cite{rodriguez/2018:CRM}.
These solution families are imposed by the boundary conditions
\begin{align} 
        \tilde{\vb*{u}} \dotproduct \vb*{n} = 0 
        \quad \text{and} \quad
        \vb*{t} \dotproduct \left(\tilde{\vb*{\sigma}} \dotproduct \vb*{n}\right) = 0
        \quad
        \text{(symmetry)}; 
        \label{eq:cond_sym} \\ 
        \tilde{\vb*{u}} \dotproduct \vb*{t} = 0 
        \quad \text{and} \quad
        \vb*{n} \dotproduct \left(\tilde{\vb*{\sigma}} \dotproduct \vb*{n}\right) = 0
        \quad
        \text{(antisymmetry)},
        \label{eq:cond_anti}
    \end{align}
in which, $\tilde{\vb*{\sigma}} = -\tilde{p}\vb*{I} + \Rey^{-1}(\nabla\tilde{\vb*{u}} + \nabla\tilde{\vb*{u}}^T)$ is the perturbation stress tensor.

The other boundary conditions for the disturbances are as follows: 

\noindent A disturbance-free inlet is defined by 
\begin{equation}
   \tilde{\vb*{u}} = \vb*{0} \quad \text{at} \quad x_2 = d_2.
\end{equation}
A zero disturbance stress condition is imposed at the outlet
\begin{equation}
\tilde{\vb*{\sigma}} \dotproduct \vb*{n} = \vb*{0} \quad \text{at} \quad x_1 = d_1.
\end{equation}
Finally, no-slip is imposed at the cylinder walls
\begin{equation}
   \tilde{\vb*{u}} = \vb*{0} \quad \text{at} \quad x_1^2 + x_2^2 = 1.
\end{equation}


\section{Results}
\label{sec:Results}

A grid convergence study is conducted for the most demanding case, $\Rey = 10^4$. For the base flow solution, the Grid Convergence Index (CGI) proposed by \citet{roache/1998}, with $F_s = 1.25$, yields $\approx 0.11 \%$ for the global enstrophy and $0.13\%$ for the recirculation length. 
For the linear stability analyses, the GCI for the growth rate ($\sigma$) and frequency ($\omega$), for the most unstable mode, are about $4.24\%$ and $0.13\%$, respectively. 
Naturally, even more accurate results are expected for $\Rey \approx 4000$, i.e., close to the critical \Rey[l] ($\Re_c$).
These quantities build confidence in the accuracy of the present results, in the context of a physical behavior analysis.

The sensitivity of the results to the truncation of the computational domain is assessed by extending the domain by $50\%$ in each direction. The change in the dominant eigenmode is negligible, with variations in $\sigma$ and $\omega$ of $0.09\%$ and $0.04\%$,
respectively.


\subsection{Base flow topology and characterization of the wake} \label{sec:steadyState} \label{sec:steady_flows}

\begin{figure}[t]
\centering
\includegraphics[width=270pt]{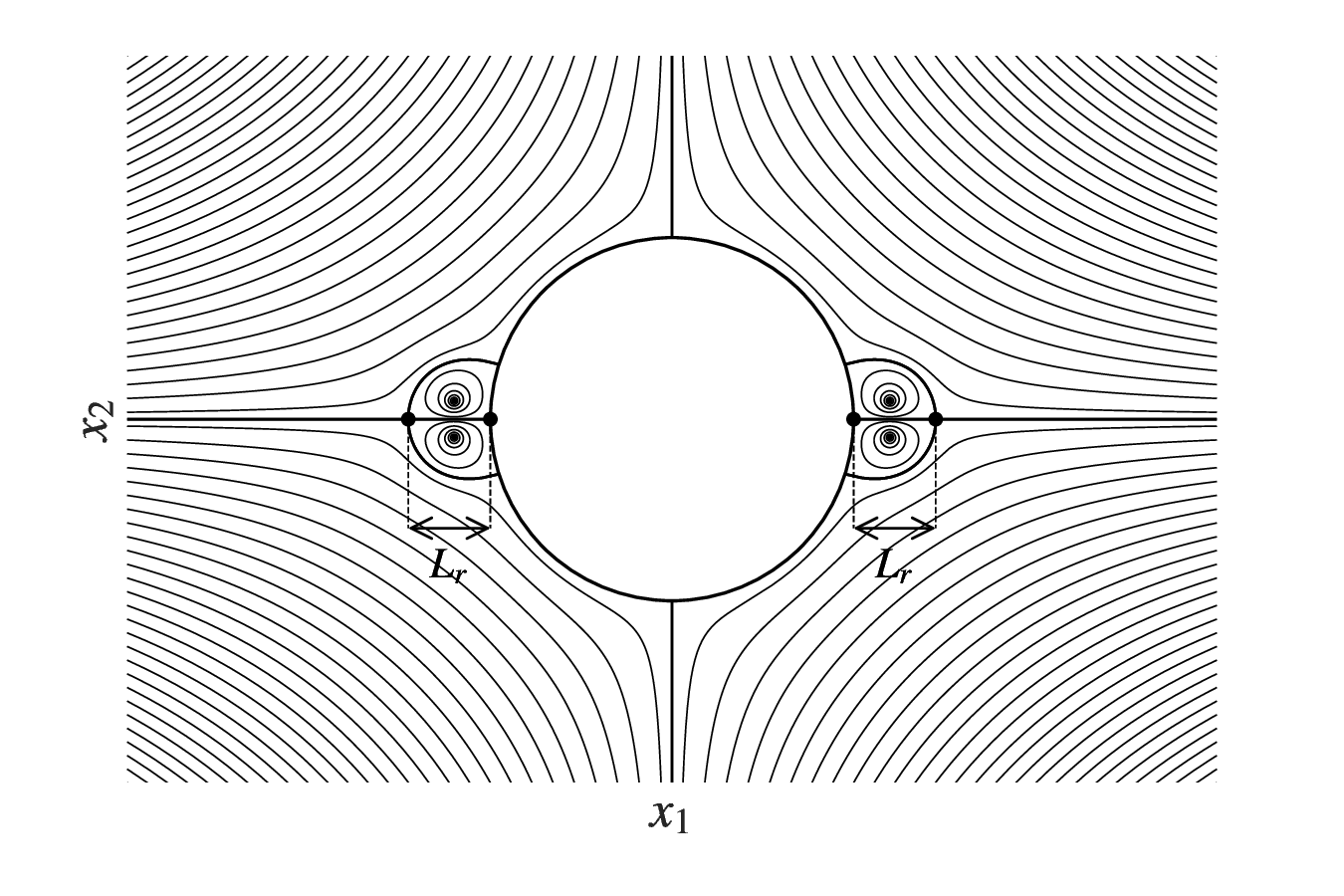}
\caption{
Flow separation with recirculation bubbles of length $L_r$.
}
\label{fig-bubbleScheme}
\end{figure}

This section studies the evolution of the steady base flow as the Reynolds number ($\Rey$) is increased. For very low values of $\Rey$, the flow is fully attached to the cylinder. The first topology bifurcation, occurring at $\Rey_{s} \approx 16.86$, is originated by flow separation from the cylinder wall and the formation of two recirculation regions, symmetric about $x_2 = 0$, as shown in Fig. \ref{fig-bubbleScheme}. The length of the recirculation region ($L_r$)  is defined as the distance between the cylinder surface and the saddle point on the horizontal symmetry axis.

\begin{figure}[t]
\centering
\begin{tabular}{cc}
(a) & (b) \\
\includegraphics[width=.49\textwidth]{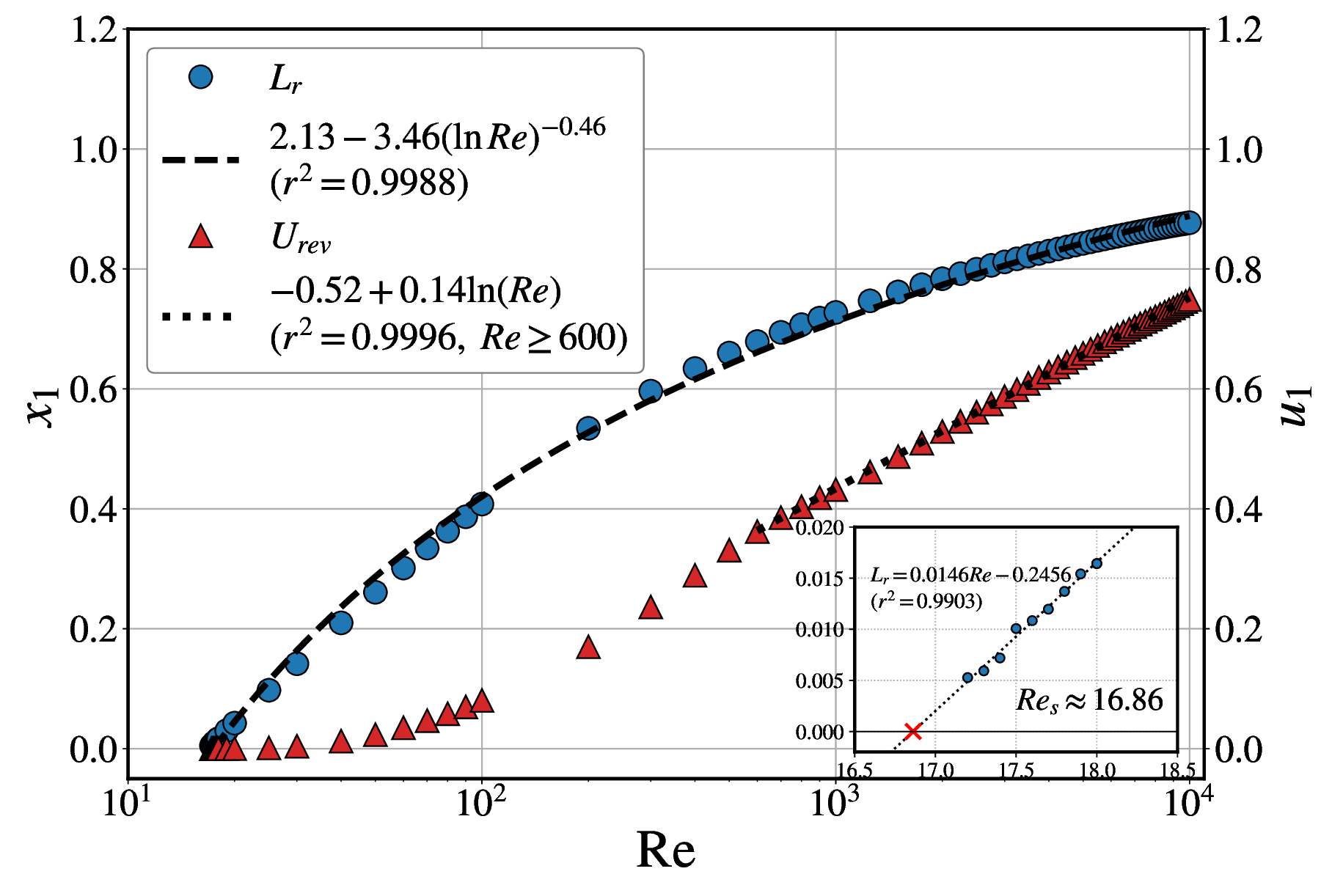} & 
\includegraphics[width=.49\textwidth]{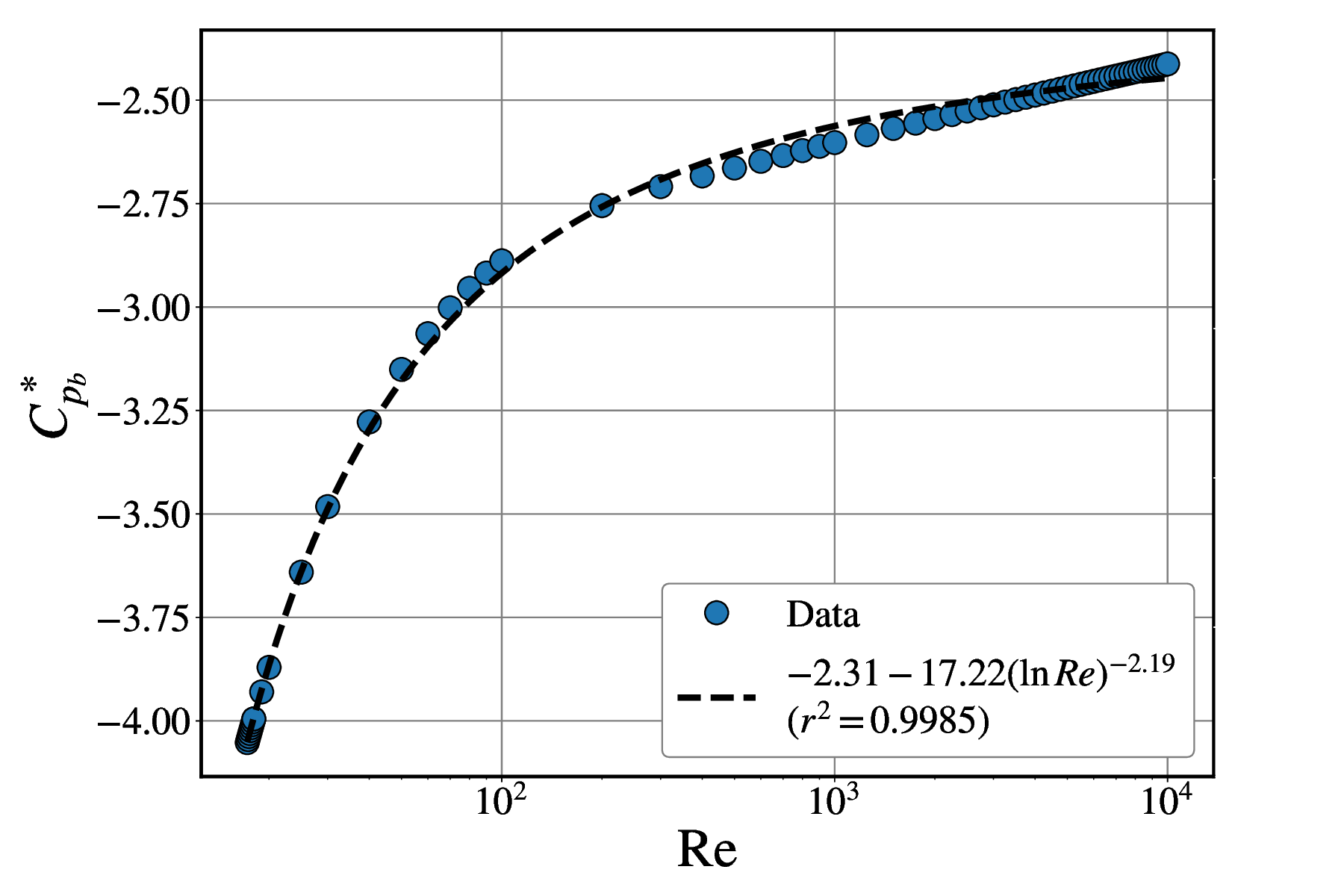} \\
(c) & (d) \\
\includegraphics[width=.49\textwidth]{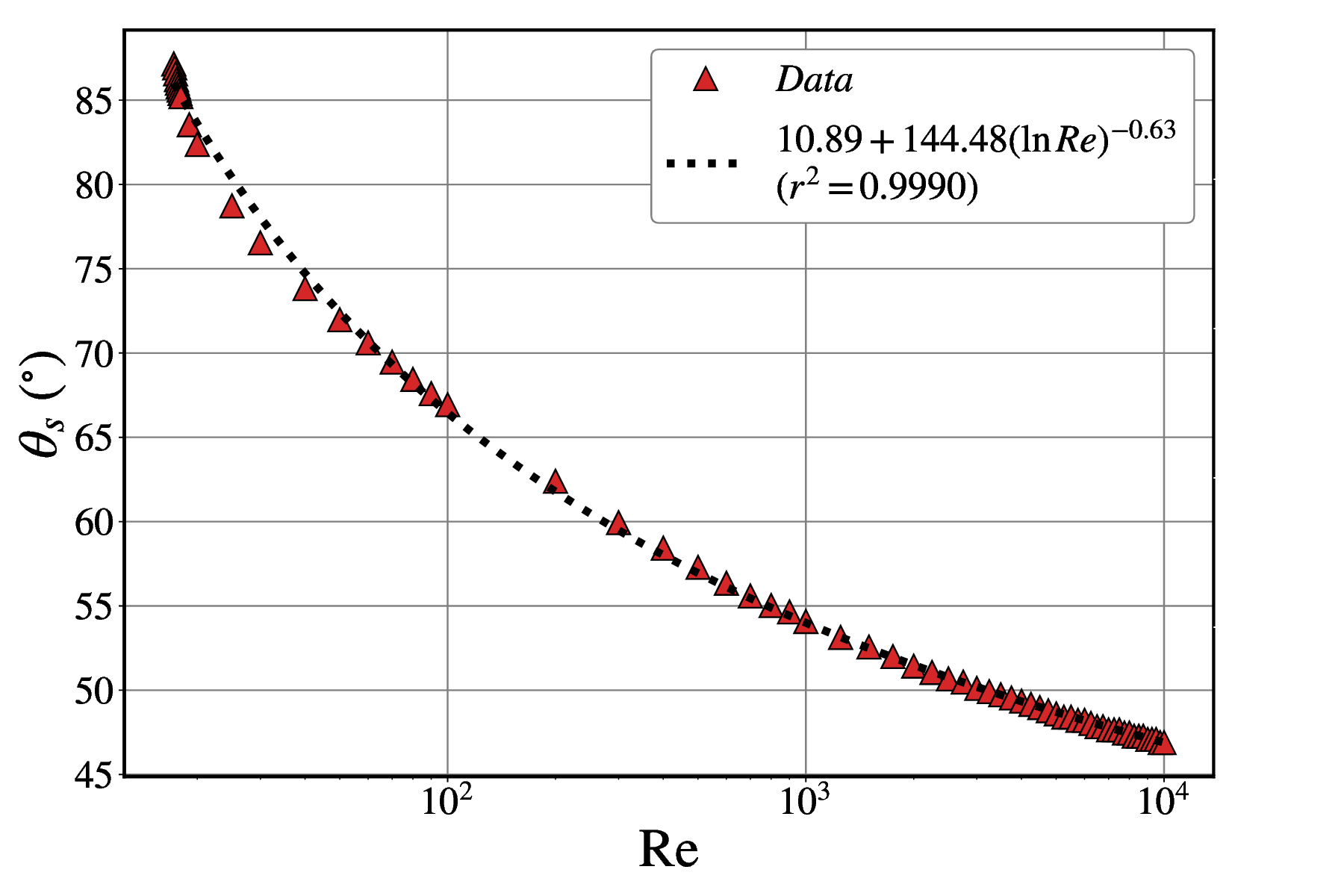} & 
\includegraphics[width=.49\textwidth]{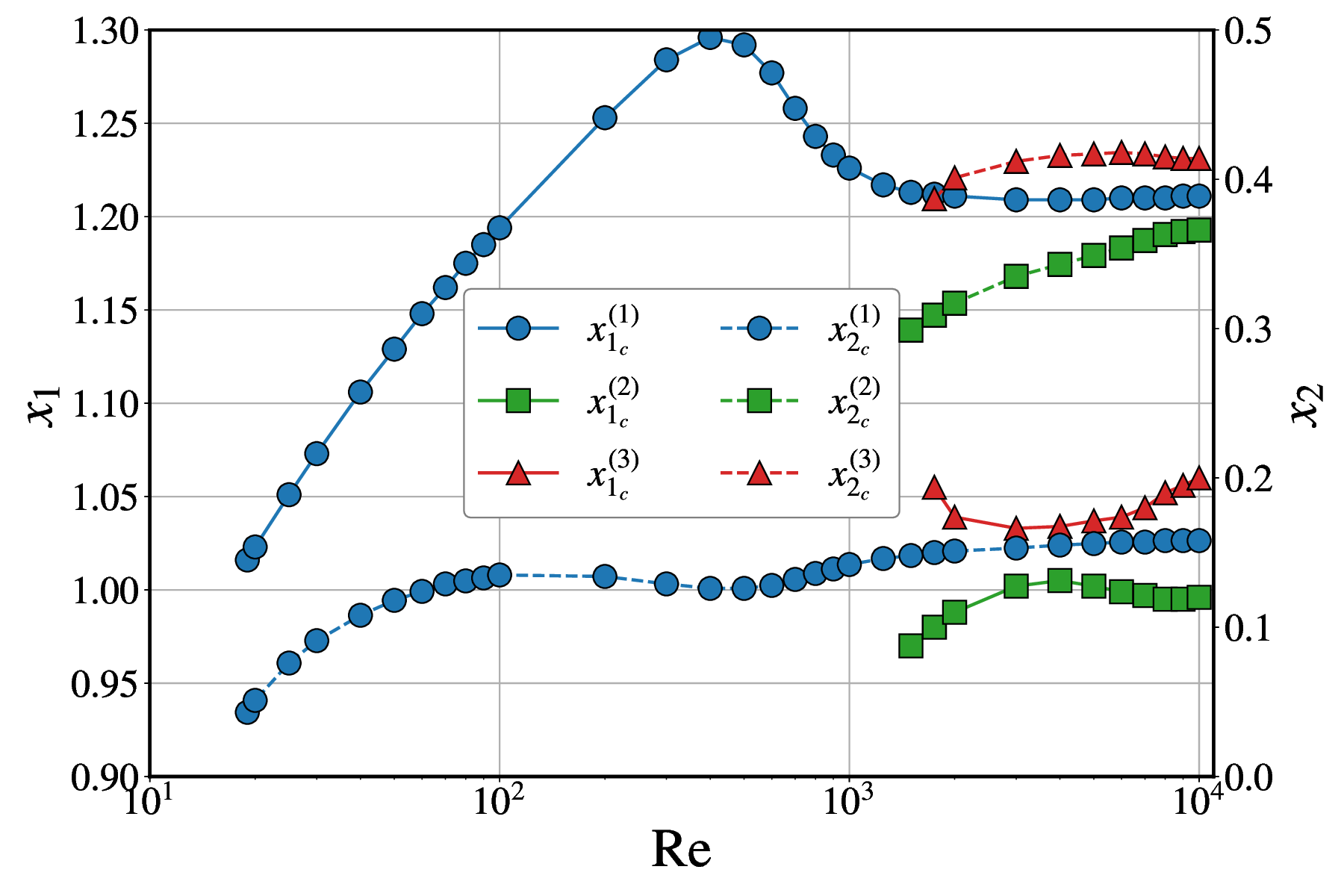}
\end{tabular}
\caption{
    The \Rey[l] dependence for: (a) recirculation bubble length, $L_r$, and maximum reversal speed, $U_{rev}$;
    (b) base pressure coefficient, $C^*_{p_b}$; 
    (c) boundary layer separation angle, $\theta_{s}$, in degrees; and
    (d) coordinates of the $i$-th (i=1,2,3) vortex center, $[x_{1_c}^{(i)}, x_{2_c}^{(i)}]^T$.
}
\label{Fig-steadyFlow}
\end{figure}

Figure \ref{Fig-steadyFlow} shows the evolution of relevant wake parameters for numerical simulations over the range of $17.2 \leq \Rey \leq 10\,000$. The lower limit $\Rey = 17.2$ corresponds to the lowest value for which separated flow was captured ($L_r \approx 0.0053$). The value $\Rey_s \approx 16.86$ is computed via linear extrapolation of $L_r$, as illustrated in the inset of Fig. \ref{Fig-steadyFlow}a. 
The upper limit at $\Rey = 10\,000$ sensibly exceeds the critical \Rey[l] ($\Rey_c \approx 4146$) for the first modal instability, as shown in Sec. \ref{sec:resultsModal}.
Other relevant parameters characterizing the base flow are: the maximum reversal speed ($U_{rev}$), defined as the absolute value of the minimum velocity on the horizontal axis; the base pressure coefficient defined as $C_{p_b}^* \coloneqq (p_b - p^*)/0.5$ for base ($p_b \coloneqq p(1,0)$) and stagnation-point ($p^* \coloneqq p(0,1)$) pressures; the separation angle with respect to the vertical axis ($\theta_s$); and the coordinates of the $i$-th vortex center, $[ x_{1_c}^{(i)}, x_{2_c}^{(i)} ]^T$, for each $i \in \{1,2,3 \}$.
The wake parameters $L_r$, $C_{p_b}^*$, and $\theta_s$ exhibit a smooth dependence with $\Rey$, which can be fitted using inverse logarithmic power laws for the range of $\Rey$ considered, as shown in Fig. \ref{Fig-steadyFlow}. For $\Rey> 600$, $U_{rev}$ can also be accurately fitted using a logarithmic function.
By using these analytic approximations, the corresponding values for $Re \rightarrow +\infty$ are estimated as $L_r^{\infty} \approx 2.13$, $C_{p_b}^{*\infty} \approx -2.31$, $\theta_s^{\infty} \approx 10.89 \degree$, and 
$U_{rev} \rightarrow 0.14 ln(\Rey)$. 
However, the onset of instability and nonlinear transition to different flow regimes can substantially modify these limit values.

\begin{figure}[t]
\centering
\begin{tabular}{cc}
(a) & (b) \\
\includegraphics[width=.4\textwidth]{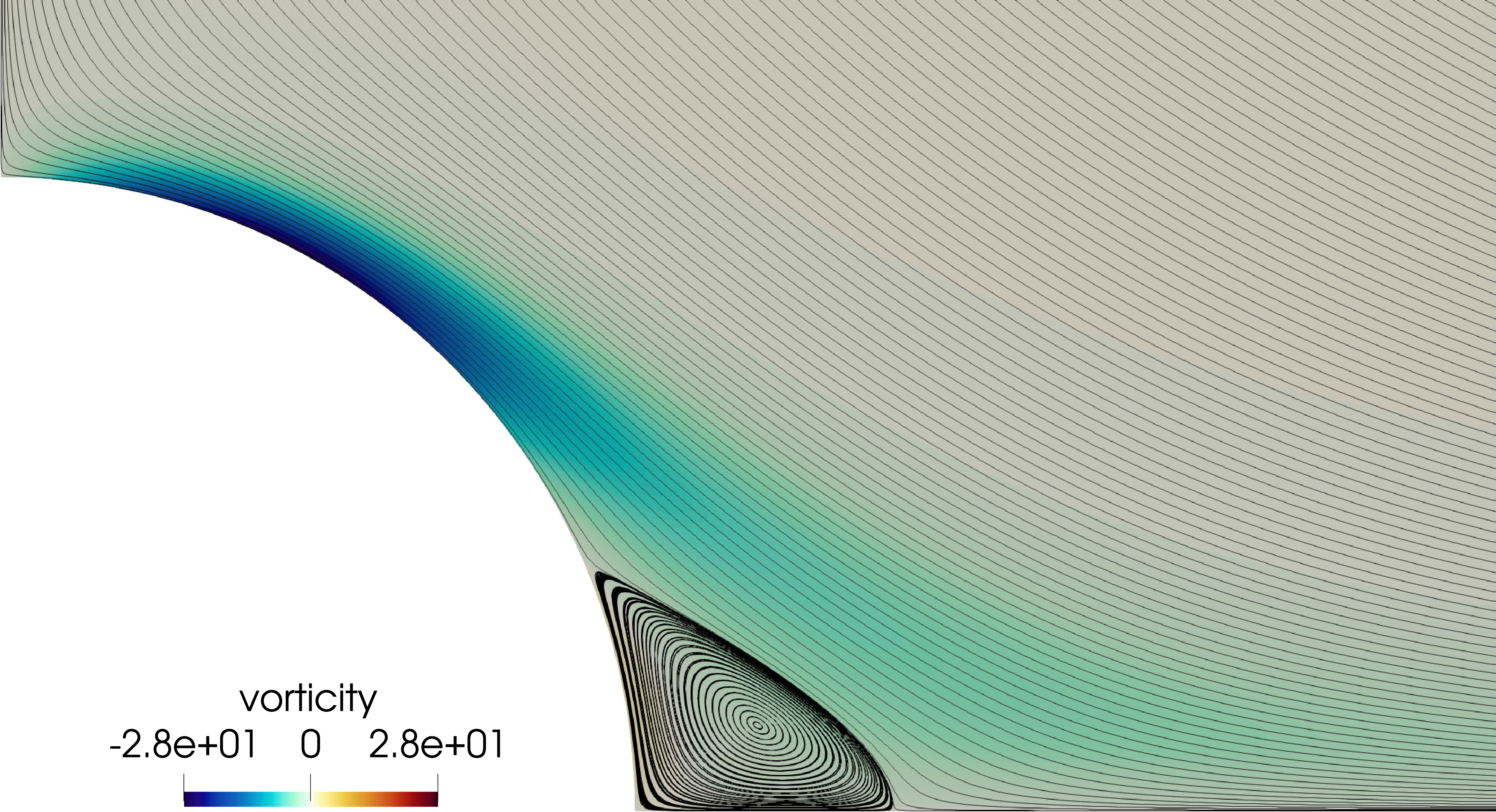} & 
\includegraphics[width=.4\textwidth]{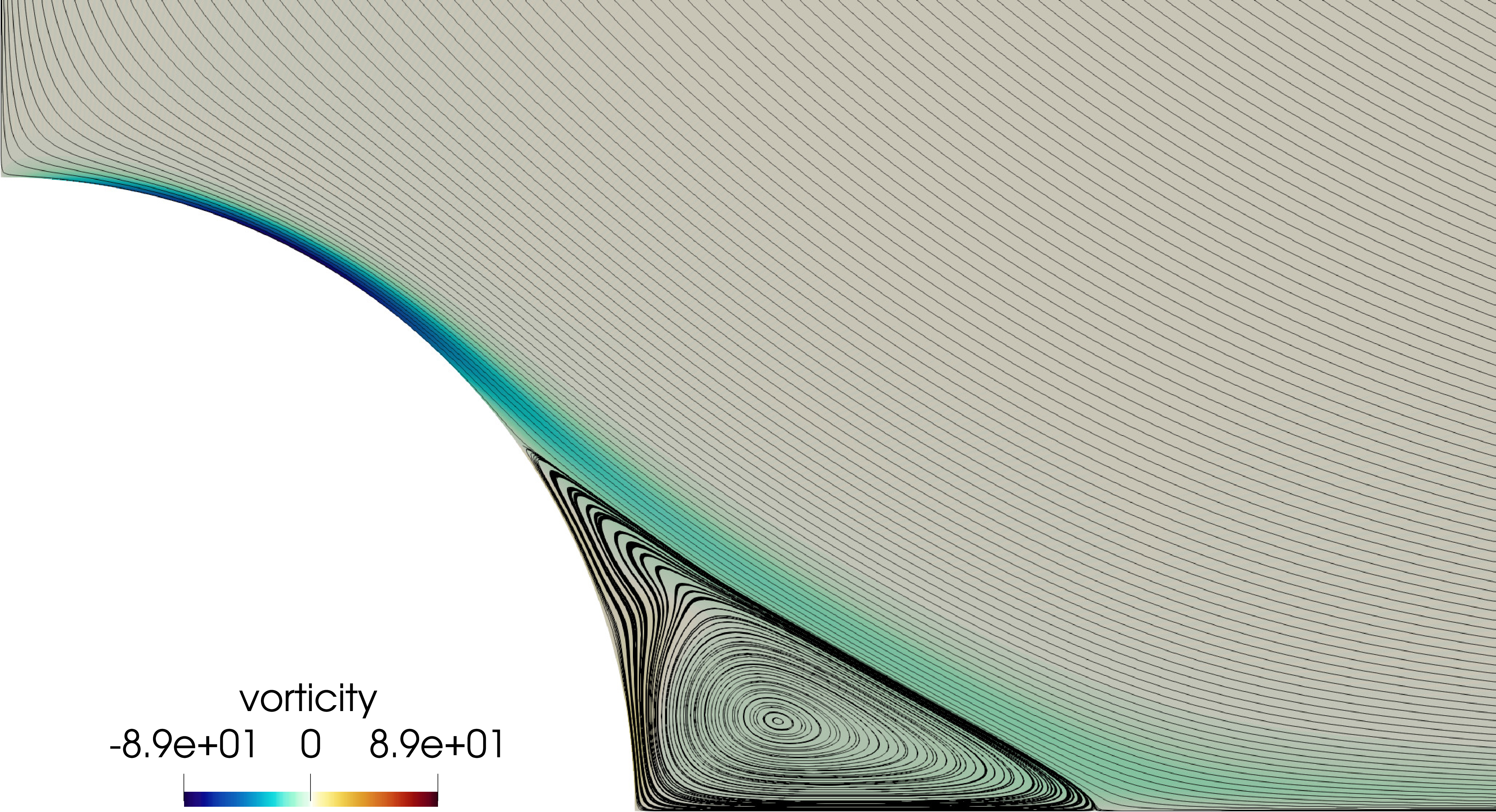} \\
(c) & (d) \\
\includegraphics[width=.4\textwidth]{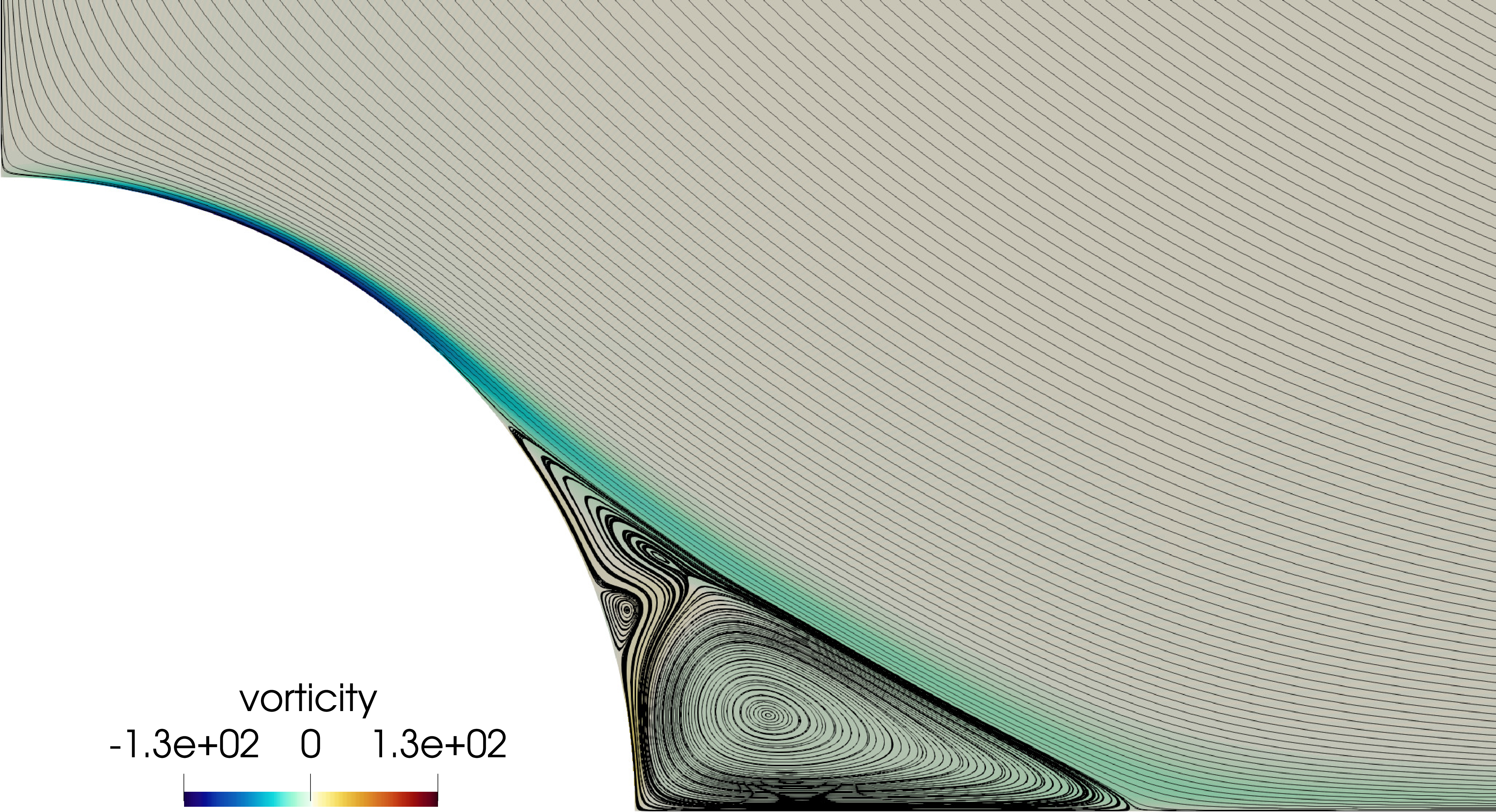} & 
\includegraphics[width=.4\textwidth]{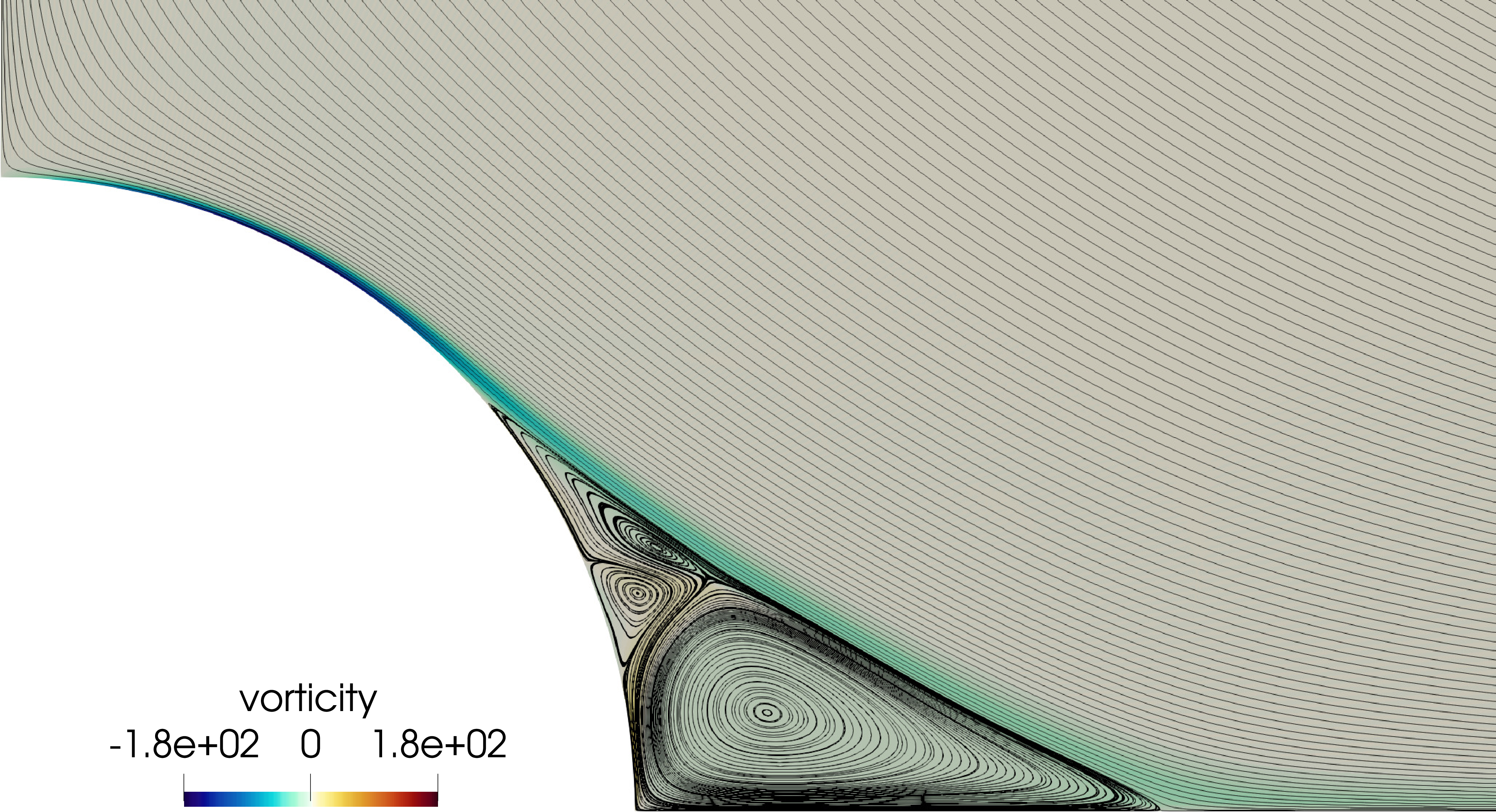} \\
\end{tabular}
\caption{
    Topology of the recirculation zone for (a) $\Rey = 100$, (b) $\Rey = 1000$, (c) $\Rey = 2000$, and (d) $\Rey = 4000$, represented by streamlines and vorticity contours.
}
\label{Fig-wakeStream}
\end{figure}

Figure \ref{Fig-wakeStream} depicts the streamlines in the recirculation region together with vorticity ($\curl{\vb*{u}}$) contours for increasing values of $\Rey$, from 100 to 4000. As the Reynolds number and $L_r$ increase, new closed recirculation regions emerge with the appearance of additional recirculation centers and saddle points, akin to Moffatt eddies \citep{Moffatt:JFM1964}; a secondary vortex appears for $\Rey \in  [1250,1500]$, and a tertiary one emerges for $\Rey \in [1500,1750]$. The coordinates of the respective recirculation centers are shown in Fig. \ref{Fig-wakeStream}.

The external strain by the counterflow exerts a strong convective acceleration on the wake region of the cylinder, inhibiting the transverse spreading of vorticity and locking the geometry of the recirculation region into a fixed shape as $\Rey$ increases.
This confinement sustains an intense mechanical pressure deficit ($C_{p_b}^*\approx -2.5$) and high shear stress, that would otherwise be alleviated by wake expansion, such as in the classical case.

\subsection{Linear stability analysis}
\label{sec:resultsModal}

\begin{figure}[t]
\centering
\begin{tabular}{cc}
(a) & (b) \\
\includegraphics[width=.49\textwidth]{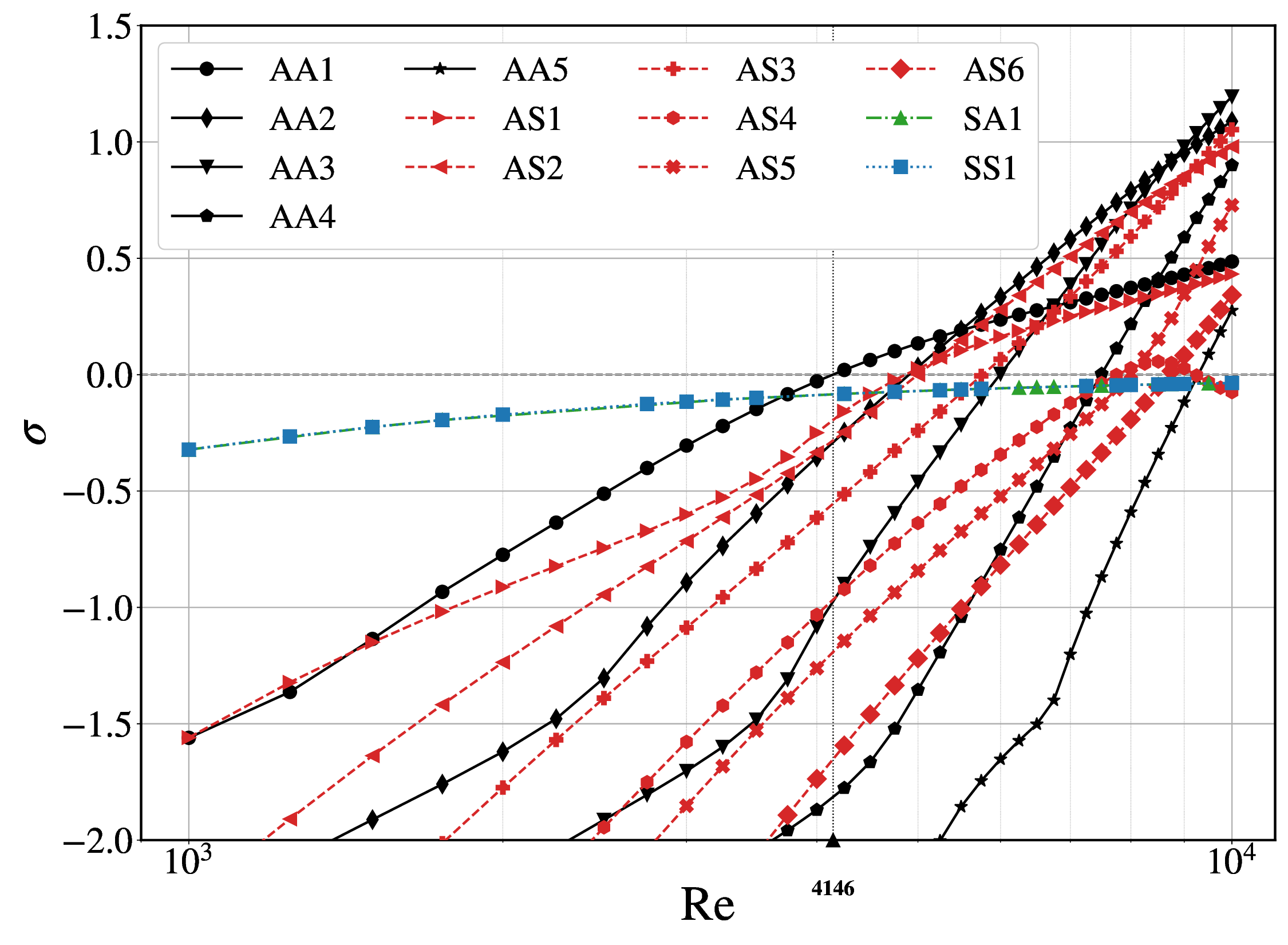} & 
\includegraphics[width=.49\textwidth]{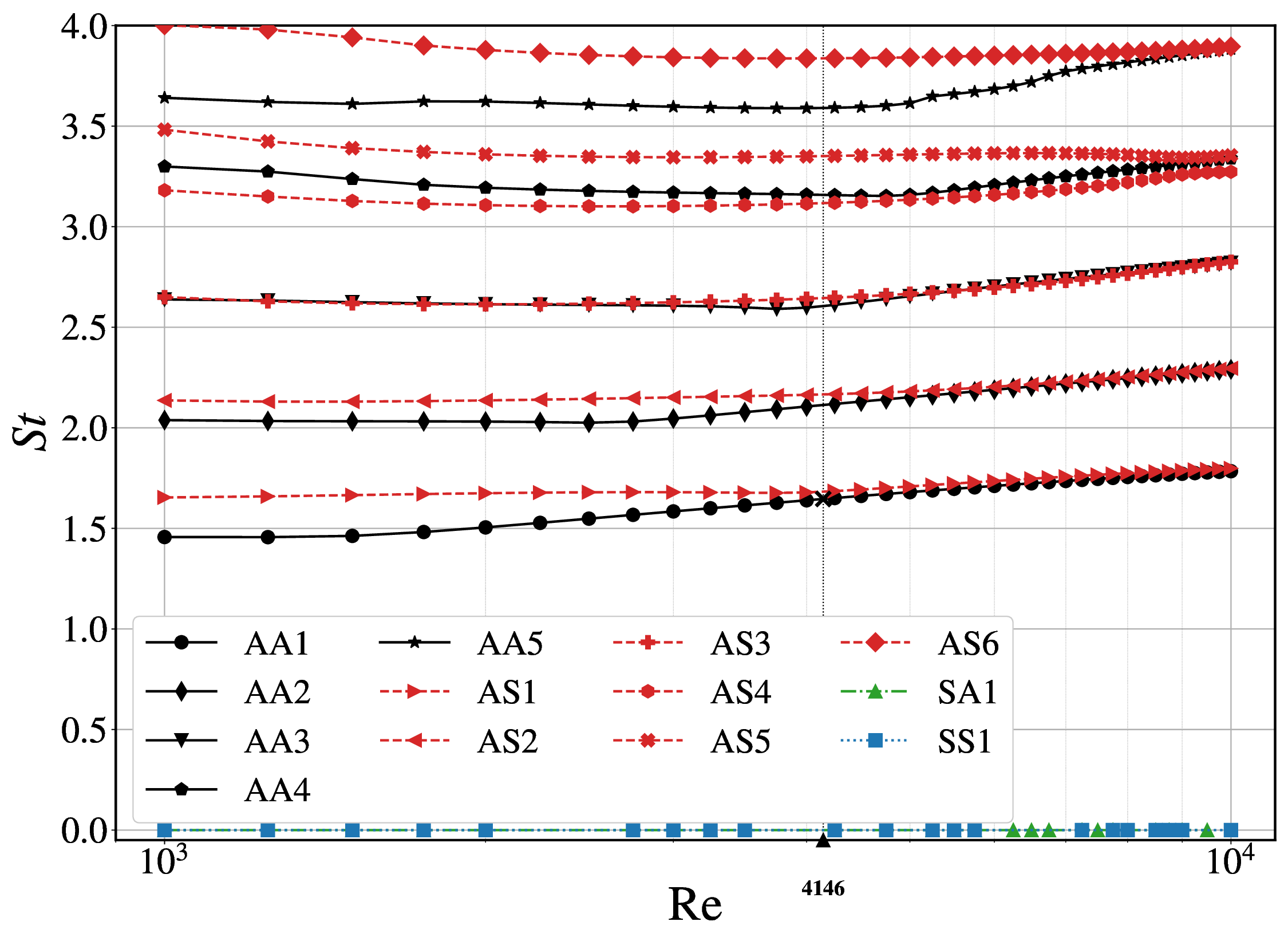}
\end{tabular}
\caption{
    Amplification rate ($\sigma$) and Strouhal number ($\St$) of the leading modes for each symmetry family (i.e., AA, AS, SA, and SS), as functions of the \Rey[l] ($\Rey$).
}
\label{Fig-LSA}
\end{figure}

Linear stability analyses are performed for $\Rey \in [1000, 10\,000]$, considering the four symmetry families as described in Sec. \ref{LSA_familiesAndBC}. The growth rate ($\sigma$) and Strouhal number ($\St \coloneqq \hat{\omega}/(2 \pi \hat{a})$ for the dominant modes of each family are shown in Fig. \ref{Fig-LSA}, as functions of the Reynolds number. 

Only modes from families AA and AS become unstable as $\Rey$ increases. The modes that are symmetric about the $x_1 = 0$ axis are stable for the entire range of Reynolds numbers analyzed. The most unstable mode for all unstable Reynolds numbers belongs to the AA family, but two switches of the dominant mode are identified. The first mode to become unstable is denominated as AA1, and becomes unstable for the critical Reynolds number $\Rey_c \approx 4146$, having a Strouhal number $\St \approx 1.65$. This mode is surpassed by mode AA2 at $\Rey \approx 5500$, which has $St \approx 2.2$ at that Reynolds number. The second switching occurs for $\Rey \approx 8500$, when mode AA3 with $St \approx 2.75$ becomes dominant. The trends of the subsequent modes from the AA family suggest that successive switches of the dominant mode would appear for even higher $\Rey$. A similar sequence of dominant unstable modes appears for the AS family, but for all conditions, the analogous AA modes are more unstable.

\begin{figure}[t]
\centering
\begin{tabular}{cc}
(a) & (b) \\
\includegraphics[width=.45\textwidth]{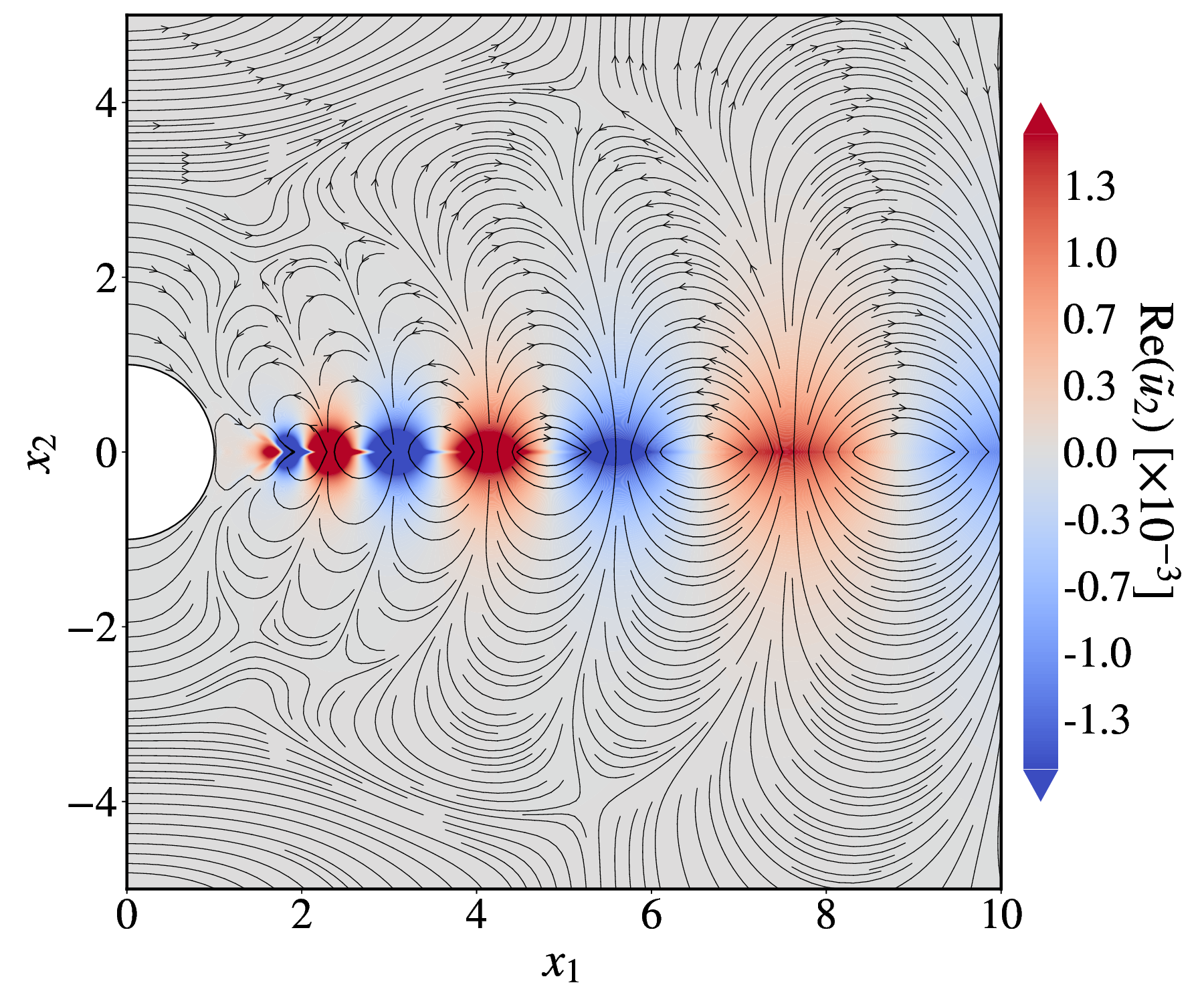} &
\includegraphics[width=.45\textwidth]{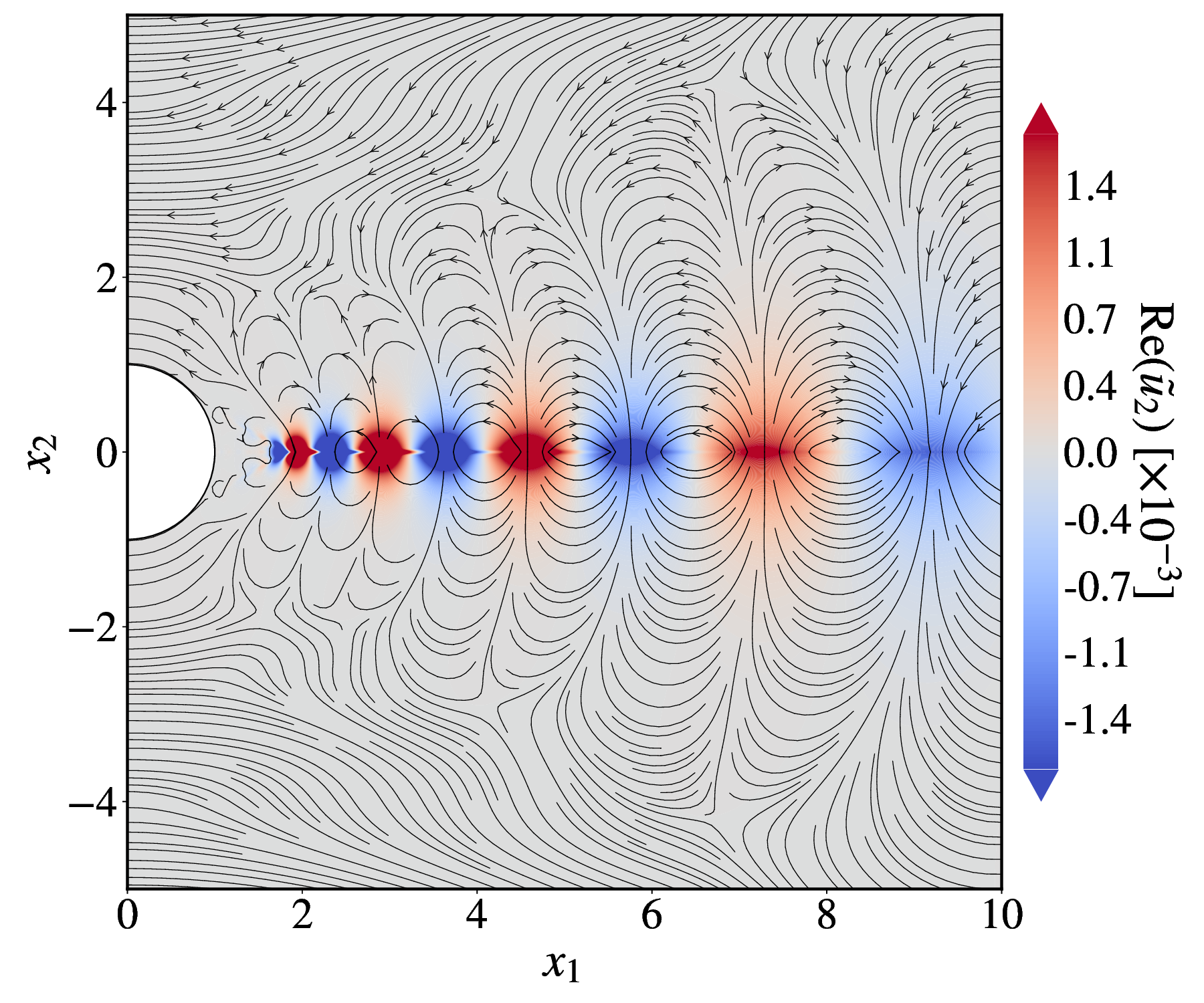} \\
\end{tabular}
\caption{
    Streamlines of flow disturbances and transverse velocity ($\tilde{u}_2$) colourmaps for modes (a) AA1 and (b) AA2 at $\Rey = 4250$ and $\Rey = 5000$, respectively.
}
\label{Fig-modes}
\end{figure}

Figure \ref{Fig-modes} shows the eigenfunctions for modes AA1 and AA2, at $\Rey = 4250$ and 5000, respectively. Streamlines of flow disturbances and transverse disturbance velocity ($\tilde{u}_2$) colormaps are shown.
Both eigenfunctions share the same qualitative features, with an alternating sequence of maxima and minima concentrated at the wake symmetry axis $x_2 = 0$, and clockwise and anticlockwise rotations. The eigenfunctions are reminiscent of those of the wake of a cylinder in a uniform flow, associated with the Hopf bifurcation that gives rise to von K\'arm\'an vortex street \citep{Zebib1987:JEM,jackson/1987:JFM}. In the present counterflow case, the apparent streamwise wavelength increases gradually along the downstream direction, following the progressive acceleration of the base flow.

\section{Conclusion}
\label{sec:Conclusions}

This work presents a characterization of the flow around a circular cylinder placed within a counterflow under two-dimensional incompressible conditions. For $\Rey < 16.86$ the flow remains attached to the cylinder wall. When $\Rey$ is increased beyond this value, a recirculation region emerges at both sides of the cylinder. The size of the recirculation region increases continuously with increasing $\Rey$, following an inverse logarithmic power of $\Rey$, combined with the emergence of additional vortices (similar to Moffatt eddies), signaling the confining effect of the counterflow. The steady flow remains stable until the critical Reynolds number $\Rey_{c} \approx 4146$, for which an unstable oscillatory mode appears, with dimensionless frequency $\St \approx 1.65$. This mode shares similarities with the one associated with the von K\'arm\'an instability in the wake of a circular cylinder immersed in a uniform flow. In the present configuration, with a cylinder immersed in a counterflow, the critical instability is anti-symmetric with respect to the two symmetry axes in the base flow. The result is a sinuous motion of the wake at each side of the cylinder oscillating in counter-phase to each other.
The sequence of bifurcations comprising the flow separation and formation of two-dimensional recirculation regions and a Hopf bifurcation towards a two-dimensional oscillatory flow is qualitatively comparable to that of the flow over a cylinder immersed in a uniform flow. However, the definition of the dimensionless parameters ($\Rey$, $\St$) is necessarily different, as the counterflow is characterized by the strain rate ($\hat{a}$) rather than a free-stream velocity ($\hat{u}_\infty$). Hence, similar values of the critical Reynolds number and the associated Strouhal number for the two configurations should not be expected.

This analysis establishes the 
foundations for new studies, including the topological transition occasioned by fluid ejection from the cylinder surface (from wake to jet), and the effects of baroclinic torque
 and burning rate oscillations on the reacting flow stability, which directly reflect on flames in a practical ambient.

\begin{acknowledgments}

This article is dedicated to the memory of Professor Amable Liñán. 

This study was financed by: the São Paulo Research Foundation (FAPESP), Brasil. Process Numbers 2021/09246-9, 2021/10689-2, and 2022/14361-4; the Conselho Nacional de Desenvolvimento  Científico e Tecnológico (CNPq) -- under grants 307922/2019-7, 161887/2021-0, and 301607/2025-7; the Coordenação de Aperfeiçoamento de Pessoal de Nível Superior -- Brasil (CAPES) -- Finance Code 001. D.R. acknowledges funding by MCIN/AEI/10.13039/501100011033 and the European Union’s FEDER (Funder ID: 10.13039/501100011033), under grant PID2024-157642MB-I00.
This research was carried out using the computational resources of the Center for Mathematical Sciences Applied to Industry (CeMEAI), funded by FAPESP (grant 2013/ 07375-0).

\end{acknowledgments}

\bibliography{ref}

\end{document}